\documentclass[aps,pra, reprint,superscriptaddress,floatfix]{revtex4-1}

\usepackage{amsmath}

\usepackage{graphicx}
\usepackage{hyperref}
\usepackage{textcomp} 

\usepackage{color}
\usepackage{soul} 

\newcommand{\TwoEightSi}{\ensuremath{^{28}\text{Si}}}
\newcommand{\TwoNineSi}{\ensuremath{^{29}\text{Si}}}
\newcommand{\ThreeZeroSi}{\ensuremath{^{30}\text{Si}}}

\newcommand{\SevenSevenSe}{\ensuremath{^{77}\text{Se}}}
\newcommand{\ThreeOneP}{\ensuremath{^{31}\text{P}}}

\newcommand{\D}{\ensuremath{\text{D}^0}}

\newcommand{\DX}{\ensuremath{\text{D}^0\text{X}}}

\begin{document}

\title{Zero field optical magnetic resonance study of phosphorus donors in 28-silicon}

\author{Kevin J. Morse}
\author{Phillip Dluhy}
\affiliation{Department of Physics, Simon Fraser University, Burnaby, British Columbia, Canada V5A 1S6}

\author{Julian Huber}
\affiliation{Vienna Center for Quantum Science and Technology, Atominstitut, TU Wien, 1040 Vienna, Austria}

\author{Jeff Z. Salvail}
\author{Kamyar Saeedi}
\affiliation{Department of Physics, Simon Fraser University, Burnaby, British Columbia, Canada V5A 1S6}

\author{Helge Riemann} 
\author{Nikolay V. Abrosimov}
\affiliation{Leibniz-Institut f\"{u}r Kristallz\"{u}chtung, 12489 Berlin, Germany}

\author{Peter Becker}
\affiliation{PTB Braunschweig, 38116 Braunschweig, Germany}

\author{Hans-Joachim Pohl}
\affiliation{VITCON Projectconsult GmbH, 07745 Jena, Germany}

\author{S. Simmons}

\author{M. L. W. Thewalt}
\email[Corresponding author: ]{thewalt@sfu.ca}
\affiliation{Department of Physics, Simon Fraser University, Burnaby, British Columbia, Canada V5A 1S6}

\date{\today}

\begin{abstract}

Donor spins in silicon are some of the most promising qubits for upcoming solid-state quantum technologies. The nuclear spins of phosphorus donors in enriched silicon have among the longest coherence times of any solid-state system as well as simultaneous qubit initialization, manipulation and readout fidelities near $\sim\!99.9\%$. Here we characterize the phosphorus in silicon system in the regime of ``zero'' magnetic field, where a singlet-triplet spin clock transition can be accessed, using laser spectroscopy and magnetic resonance methods. We show the system can be optically hyperpolarized and has $\sim\!10$\,s Hahn echo coherence times, even at Earth's magnetic field and below.

\end{abstract}

\maketitle

\section{Introduction}

The electron and nuclear spins of donor impurities in silicon continue to be attractive qubits for upcoming quantum devices.  The shallow phosphorus donor remains the most frequently studied, due to its ubiquity, long coherence times \cite{Steger:2012,Saeedi:2013}, and demonstrated performance in single and multiple-qubit devices \cite{Pla:2012,Pla:2013,Dehollain:2014,Dehollain:2015,broome:2017,Harvey-Collard:2017}. Most of the published literature for this system is in the relatively high magnetic field regime, using either inductively-detected magnetic resonance on ensembles of donors \cite{Feher:1959,tyryshkin:2012,Wolfowicz:2012,franke:2015}, typically at an applied field ($B_0$) of $\sim\!0.32$\,T,  or single-donor initialization and readout via spin-dependent tunnelling \cite{Morello:2010}, which requires an applied field of greater than 1~T.  These hyperfine-coupled donor spin systems also exhibit what are called ``clock transitions'' \cite{Wolfowicz:2013} where at a specific $B_0$ the frequency of a particular spin flip transition has zero first-order dependence on $B_0$. This results in an insensitivity to fluctuations in $B_0$, which can result in an orders of magnitude increase in coherence time \cite{Steger:2012}.  Phosphorus has a nuclear spin $I=1/2$, and therefore for $B_0>0$ has only a single NMR-like clock transition at $B_0 \approx 845$\,G \cite{Steger:2011jap}.  The other shallow donors in silicon, having $I > 1/2$, exhibit both NMR-like and EPR-like clock transitions, but these are all realized at fairly large magnetic fields \cite{Wolfowicz:2013}.

All shallow donors in silicon have another clock transition at $B_0=0$, which has so far not been investigated due to the very small thermal polarizations available under these conditions.  Here we show that the optical hyperpolarization and readout methods, using the shallow donor bound exciton (\DX{}) optical transitions already demonstrated \cite{yang:2009} for the shallow donors at higher $B_0$, can also produce rapid and large hyperpolarization at $B_0 \approx 0$, and rapid read out of the hyperfine populations, allowing for magnetic resonance measurements at or near $B_0=0$.  We note that similar magnetic resonance measurements have recently been demonstrated for the deep chalcogen donor $\SevenSevenSe^+$, but using a different method of optical hyperpolarization and readout \cite{Morse:2017}.  We further note that while spin-dependent recombination methods have been demonstrated to enable magnetic resonance measurements on donors in silicon at small $B_0$ \cite{dreher:2015,Mortemousque:2016}, they have not been shown to work all the way down to zero field.  In addition to eliminating the need for a large and homogeneous magnetic field, these donor spin clock transitions near $B_0=0$ could also simplify the realization of hybrid donor spin/superconducting resonator coupling schemes \cite{Bienfait:2016}.

\section{Phosphorus Donor System}

The quantum system used in this work is the electron and nuclear spins of the phosphorus donor (\ThreeOneP{}) in isotopically enriched \TwoEightSi{}. Naturally occurring silicon consists of three isotopes, \TwoEightSi{}, \TwoNineSi{}, and \ThreeZeroSi{}, and by essentially removing the other two isotopes we remove spatial variations in the local band gap energy, as well as the nuclear spins of the \TwoNineSi{}, resulting in a so called ``semiconductor vacuum'' \cite{Steger:2012}.  The system is then analogous to a hydrogen atom in a vacuum \cite{Cardona:2005}. For this study only the ground electronic state of the neutral phosphorus donor is relevant, with a binding energy of $\sim\!45$\,meV \cite{aggarwal:1965}.

\begin{figure}[htbp]
\centering
\includegraphics[width=\columnwidth]{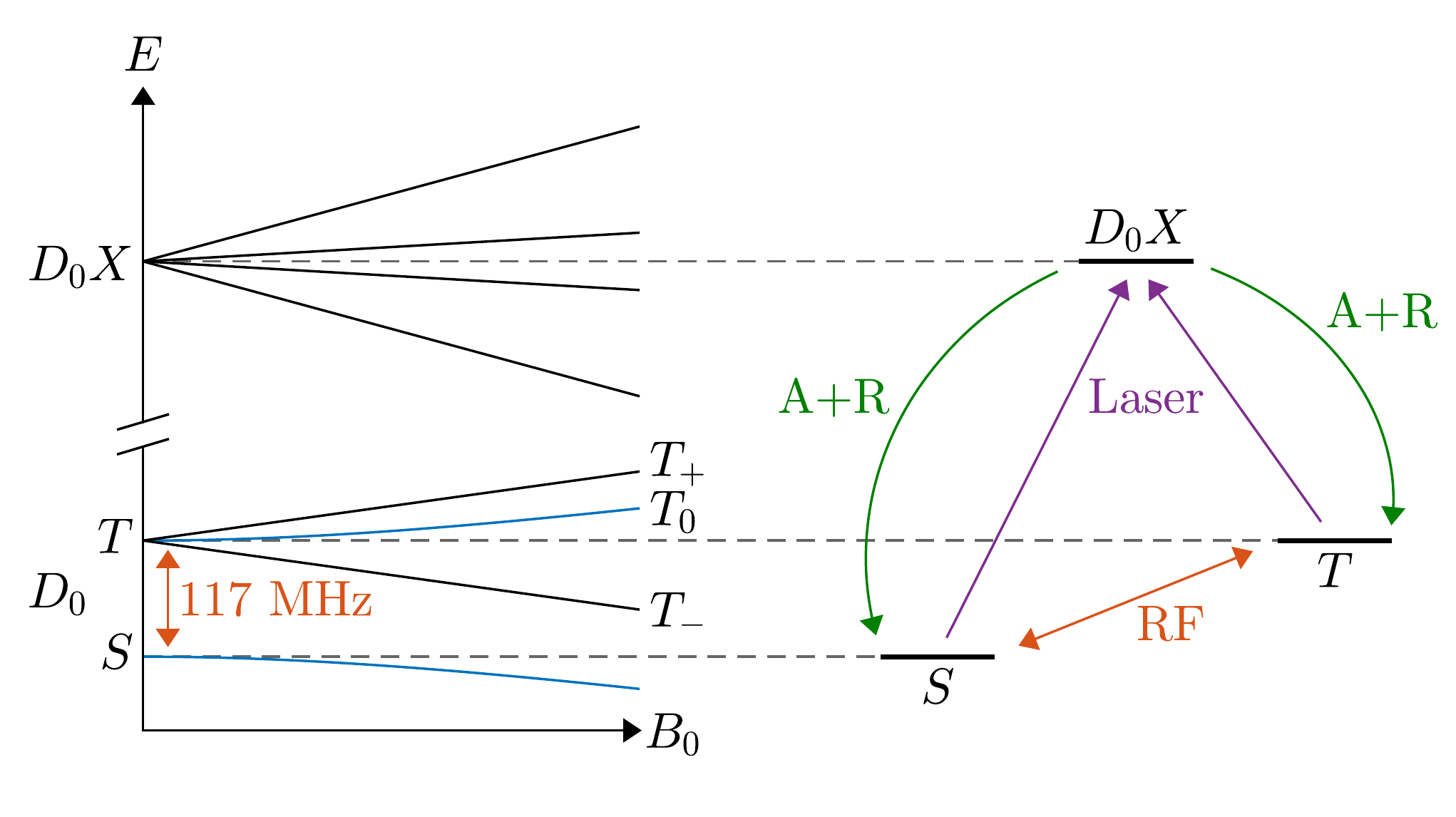}
\caption{\label{fig:energy_levels}Energy levels of the \ThreeOneP{} neutral donor (\D{}) Zeeman states from 0 to 5 mT (left) and the transitions used in this work (right). The 117\,MHz radio frequency (RF) transition between the singlet and triplet states is shown in red, the two optical transitions from the singlet and triplet to the donor bound exciton (\DX{}) in purple, and the decay from \DX{} to either the singlet ($S$) or triplet ($T$) state of \D{} through the Auger process followed by electron recapture ($\text{A}+\text{R}$) is shown in green.}
\end{figure}

Considering only the contact hyperfine interaction between the donor electron and nuclear spin, the spin Hamiltonian for the neutral donor (D$^0$) ground state in an applied magnetic field $B_0\hat{z}$ can be written,
\begin{equation} \label{eq:hyperfinehamiltonian}
\mathcal{H}_{\text{D}^0} =  \gamma_S B_0 S_z -  \gamma_I B_0 I_{z} + A \vec{S} \cdot \vec{I},
\end{equation}
where $A$ is the hyperfine constant and $\vec{S}$ and $\vec{I}$ are the spin operators of the electron and nuclear spin respectively. The constants $\gamma_S$ and $\gamma_I$ are defined as follows,
\begin{equation*}
\gamma_S = \frac{g_e \mu_B}{h},\quad \gamma_I = \frac{g_n \mu_N}{h},
\end{equation*}
where $g_e$ and $g_n$ are the electron and nuclear $g$-factor, $\mu_B$ and $\mu_N$ the Bohr and nuclear magneton, and $h$ the Planck constant. 
We divide by $h$ in order to work in frequency units where for \ThreeOneP{}, $A \approx 117.53$\,MHz, $\gamma_S = 27.972$\,MHz/mT and $\gamma_I = 17.251$\,kHz/mT. The eigenvalues of \eqref{eq:hyperfinehamiltonian} are given by the Breit-Rabi formula and plotted in Fig.~\ref{fig:energy_levels}.

For $B_0=0$, these eigenvalues reduce to two levels, with total spin $F=1$ and $F=0$, separated by $A$.  For $B_0>0$, such states split into $2F+1$ Zeeman components, giving a triplet $T$ with three $F=1$ levels $\left(T_-, T_0, T_+\right)$ and a singlet $S$ with $F=0$.

By applying a radiofrequency (RF) magnetic field, $B_1$, at a specific frequency we can drive transitions between the singlet and the triplet states. The $S \rightarrow T_{\pm}$ ($S \rightarrow T_{0}$) is an allowed transition when $B_1$ has a component perpendicular (parallel) to $B_0$, and this remains true in the limit where $B_0=0$ \cite{Hayden:1995}. At the fields used in this work ($B_0 < 25$\,\textmu{}T) the $S \to T_0$ transition frequency is essentially constant, since it is a clock transition, while the $S \to T_{\pm}$ transition frequencies vary linearly with $B_0$.  Near zero field the $S \to T_0$ transition should be much less sensitive to inhomogeneities and noise in $B_0$ than the $S \to T_{\pm}$ transitions.

Studies at nonzero $B_0$ have shown that the removal of inhomogeneous isotope broadening, realized by using highly enriched \TwoEightSi{}, makes it possible to resolve the hyperfine-split \D{} ground state components in the absorption spectrum of the donor bound exciton (\DX{}). Optical transitions between these components can also be used to hyperpolarize the spin system and to measure the populations in the various hyperfine states \cite{yang:2009}. These same methods can be applied near $B_0=0$, as shown on the right hand side of Fig.~\ref{fig:energy_levels}.  The resonant laser radiation selectively promotes donors in either the $S$ or $T$ states to \DX{}, and the predominantly nonradiative Auger decay of the \DX{}, followed by recapture of the free electron to either the $S$ or $T$ state drives the hyperpolarization \cite{yang:2009}.  The Auger electrons can also be used to measure a photoconductive signal which is proportional to the number of \DX{} generated, and thus to the \D{} population in the state being pumped \cite{Steger:2012}.  Note that while the optical \DX{} transition can resolve the $S$ and $T$ states, it cannot at these low fields resolve the $T_0$, $T_+$ and $T_-$ states.  These can only be probed under the present conditions by using magnetic resonance.

\section{Sample and Apparatus}

\subsection{Sample}

The sample studied for this research was a small $5.0 \times 4.7 \times 1.7$\,mm piece cut from a slice of the Avogadro crystal (Si28-10Pr11.02.2). This crystal was grown by the Leibniz Institute for Crystal Growth (Leibniz-Institut f\"{u}r Kristallz\"{u}chtung) as part of the Avogadro project and was enriched to $99.995\%$ $^{28}$Si \cite{Becker:2010}. After being cut to size, the sample was etched in a $10:1$ mixture of $\text{HNO}_3 : \text{HF}$ to remove surface damage. This step was important as surface damage causes strain which leads to splitting and broadening of the spectral lines.

Impurity concentrations were determined using photoluminescence spectroscopy to be $\sim\!5\times10^{11}\,\textrm{cm}^{-3}$ phosphorus and $\sim\!5\times10^{13}\,\textrm{cm}^{-3}$ boron. The sample was also found to contain $<1\times10^{14}\,\text{cm}^{-3}$ oxygen and $<5\times10^{14}\,\text{cm}^{-3}$ carbon.

\subsection{Cryogenic Assembly}

The sample was mounted in a 3D printed sample holder designed to locate the sample between two pieces of copper foil (used for the photoconductive readout scheme) in a strain-free manner. This sample holder was mounted in the centre of a plastic Helmholtz coil form. The Helmholtz coil was tuned to approximately 117\,MHz and impedance matched to a 50\,$\Omega$ transmission line. 

This assembly was inserted into a non-magnetic stainless steel immersion dewar made by Janis Research. The dewar used was specifically chosen as it had never been placed in a magnetic field. This was important as preliminary testing showed remnant fields existed in other similar dewars that had previously been used in high magnetic fields. 

The tail of the dewar was shielded by wrapping several layers of high permeability 80\% nickel alloy foil around it. For further shielding the entire bottom section of the cryostat had two layers of woven magnetic shielding material wrapped around it. After the shielding was added the field was measured to be around 4\,\textmu{}T. In order to get to even lower fields, three coil pairs ($\hat{x}$, $\hat{y}$, and $\hat{z}$) were wound around the tail of the dewar but within the shielding. The dewar tail had fused silica windows and a single small hole was cut into each layer of shielding to allow the laser light to reach the sample. 

\subsection{Optical Excitation}

Most of the measurements were performed using a Yb-doped single-frequency tunable fibre laser and a 1047\,nm diode pumped laser. The fibre laser was used for initialization and readout while the 1047\,nm diode laser was used to provide weak above-gap excitation, which is needed to create \D{} since the sample is p-type.  Above-gap excitation produces free carriers which photoneutralize the ionized donors and acceptors, reducing random electric fields within the sample, which in turn sharpens the optical \DX{} transitions. 

An additional external-cavity tunable diode laser was employed as a pump laser for the two scans in Fig.~\ref{fig:optical-scan}. Both the Yb-doped fibre laser and the tunable diode laser are locked and scanned with respect to a stable reference cavity, which was itself locked to a frequency stabilized HeNe laser. This provided long term laser frequency stability and repeatability of a few MHz. 

Both lasers were connected to Yb-doped fibre amplifiers as well as an optical switch which allowed either laser wavelength to be measured by a Bristol 621 wavelength meter. The outputs of the laser amplifiers were collimated into free space beams, passed through shutters, and aligned to the same beam path.

\subsection{Readout Circuit}

The setup used lock-in detection of an AC signal capacitively coupled through the sample impedance. The driving signal was a sine-wave with frequency around 117\,kHz and amplitude of $\sim\!20\,\textrm{V}_{\textrm{pp}}$. The sample was mounted between two copper electrodes, which served as the plates of a parallel plate capacitor, generating a field $\vec{E}$ across the sample. In parallel with the sample was a phase shifting (PM) and amplitude attenuating (AM) circuit in series with a 12\,pF capacitor. This allowed the signal into the low noise preamplifier (LNA) to be nulled (minimized) for a specific experimental condition, e.g.~a sample in the dark or a sample in steady-state under illumination by the read-out laser. An oscilloscope in XY mode assisted with the setting of this null condition.  A lock-in measured this out-of-null signal at 117\,kHz to generate a result proportional to the change in sample impedance produced by the Auger carriers.

\section{Results}

\begin{figure}[htbp]
\centering
\includegraphics[width=\columnwidth]{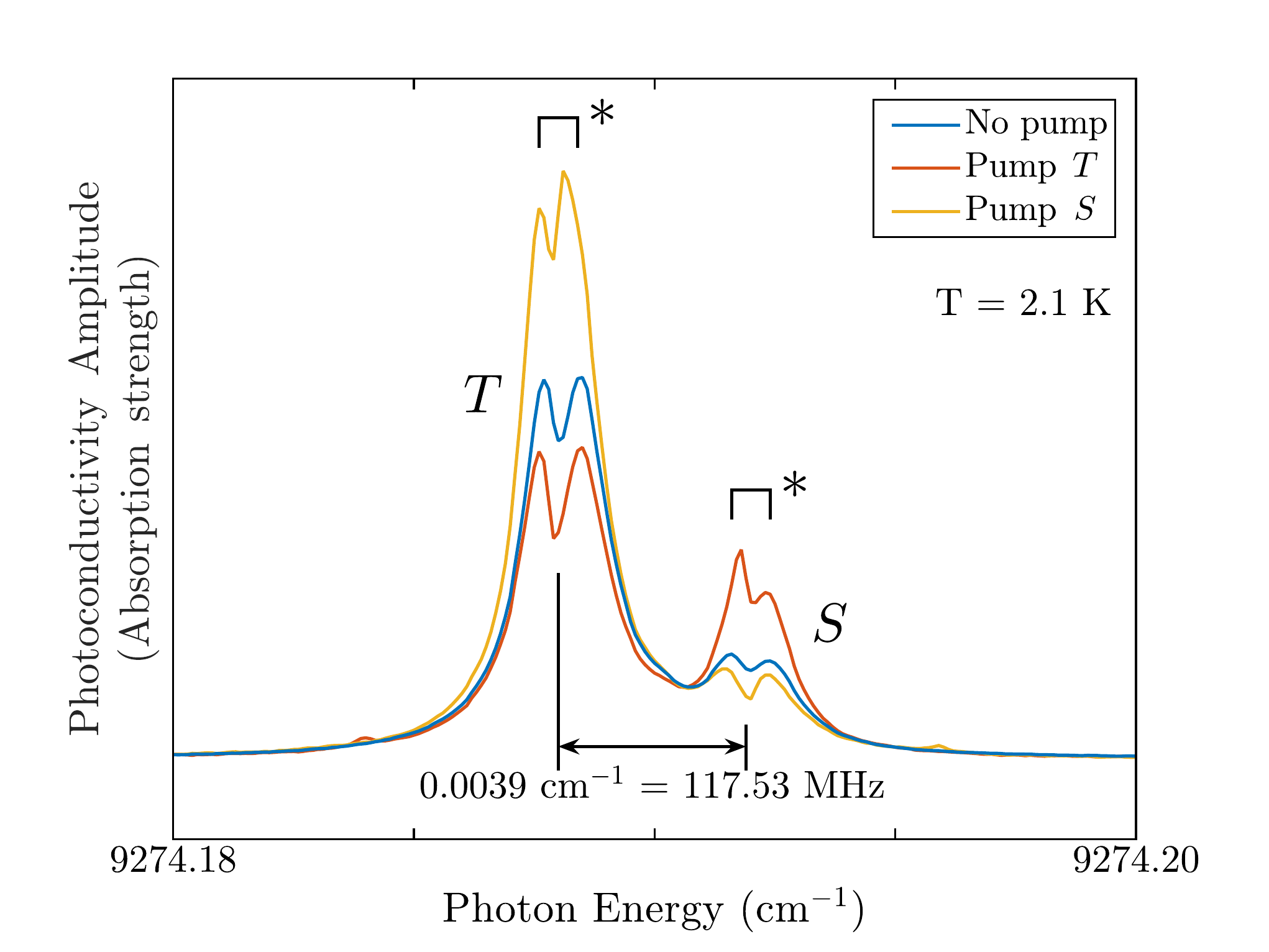}
\caption[Optical spectra -- ``zero''-field]{``Zero'' field optical spectra. Conditions common to all spectra: 20\,mW probe, 400\,mW pump, 70\,\textmu{}W above-gap excitation. The blue trace is with the pump laser turned off, the orange trace is with the pump laser on the triplet at $9274.188\,\text{cm}^{-1}$, and the yellow trace is with the pump laser on the singlet at $9274.192\,\text{cm}^{-1}$. The asterisks ($\ast$) indicate a splitting of $0.0008\,\text{cm}^{-1}$ in each line that arise from random fields due to the imperfect isotopic enrichment \cite{yang:2006}, which in natural silicon at $B_0=0$ causes related splitting of the acceptor ground state \cite{Karaiskaj:2002,Karaiskaj:2003}. \label{fig:optical-scan}
}
\end{figure}

To verify the readout mechanism, optical spectra of the transitions were recorded by scanning the fibre laser across the transitions and recording the output of the lock-in amplifier. The same scan was then repeated with the addition of a high powered pump laser tuned to either of the lines. Figure~\ref{fig:optical-scan} shows the spectra with no pump laser, the pump laser on $T$, and the pump laser on $S$. From the figure we see that not only is hyperpolarization achievable by pumping either of the lines but that significant population can be moved from one state to the other.  Note that the results shown in Fig.~\ref{fig:optical-scan} are not intended to indicate the maximum possible hyperpolarization which can be generated, since a considerable amount of above-gap excitation was used while collecting these spectra to prevent the read-out laser itself from saturating the system (the above-gap excitation counteracts the hyperpolarization by acting to randomize the hyperfine populations).

\begin{figure}[htbp]
\centering
\includegraphics[width=\columnwidth]{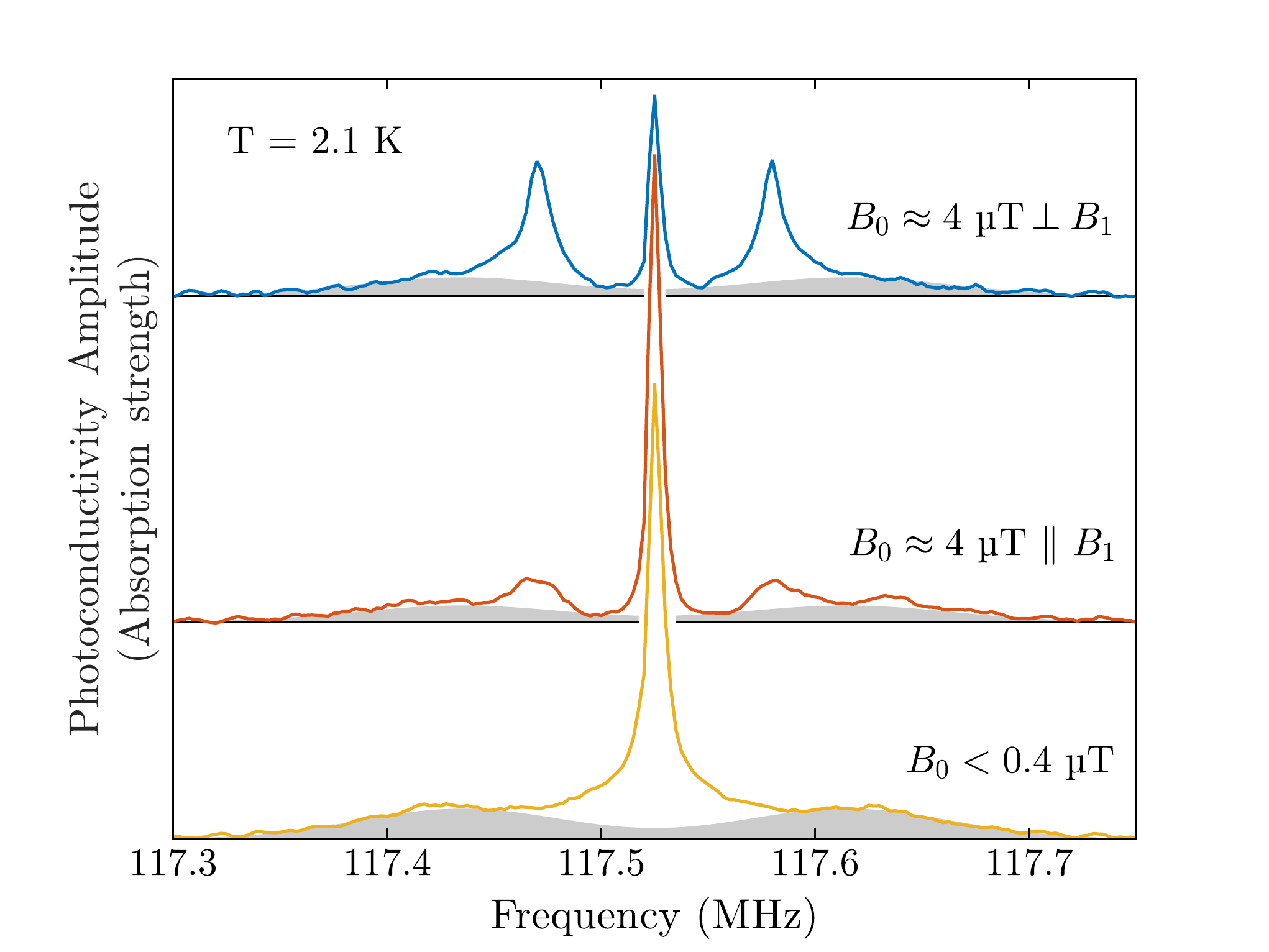}
\caption[MR spectra -- small fields]{Small field magnetic resonance spectra. Conditions common to all spectra: 130\,mW pump/probe at $9274.1888\,\text{cm}^{-1}$ and 7\,\textmu{}W above-gap excitation. The gray shaded regions are discussed in the text. \label{fig:rf_small_field}}
\end{figure}

After confirming the system could be hyperpolarized, the probe laser was set on the $T$ line and RF was applied to the $B_1$ coil. The RF frequency was then scanned across the expected transition frequencies and the photoconductive signal was recorded (See Fig.~\ref{fig:rf_small_field}). By measuring the frequency difference between the peaks of the $S \rightarrow T_{+}$ and $S \rightarrow T_{-}$ lines, the remnant $B_0$ field was estimated to be around 4\,\textmu{}T with no current through the three active shielding coil pairs.

Current was then applied to all three coil pairs and adjusted to cancel out as much of the remaining $B_0$ as possible. This result is shown in the bottom-most trace of Fig.~\ref{fig:rf_small_field}. The $B_0$ in this configuration was estimated to be less than 0.4\,\textmu{}T.  Currents through the coils $\perp B_1$ and $\parallel B_1$ were separately adjusted until the measured splitting was approximately equal to 4\,\textmu{}T. These results are shown in the top and middle traces of Fig.~\ref{fig:rf_small_field} and allow comparison between the strength of the $S \to T_{\pm}$ and $S \to T_0$ lines in two different $B_0$ configurations. 

The $S \to T_{\pm}$ transitions are much stronger when $B_0\perp B_1$ while the $S \to T_0$ is stronger when $B_0 \parallel B_1$. Also of note are the broad Gaussian areas (in gray) on either side of the $S \to T_0$ transition. The splitting between them seems to remain constant regardless of the $B_0$ applied. Calculating the field that would cause such a splitting gives a $B \approx 6$\,\textmu{}T. That these broad lines don't shift with small changes in $B_0$ suggests that they are due to internal fields that can't be cancelled out by the external cancellation coils, but which influence only a subset of the donors.

After performing magnetic resonance scans at each different $B_0$ configuration we moved on to pulsed measurements. For each $B_0$ and transition, Rabi and Ramsey measurements were used to determine the $\pi$-pulse length and transition frequency.  Pulsed measurements were performed by optically hyperpolarizing the system, blocking the laser excitation while the RF was applied, and then reapplying the laser and measuring the impedance transient as the laser repolarized the system to its initial state.

Hahn echo measurements were then performed using the pulse sequence $\pm\pi/2 : \tau : \pi : \tau : \pi/2$ with the first pulse alternating between $+\pi/2$ and $-\pi/2$ (phase cycled by 180 degrees) and the length of $\tau$ increasing after every two sequences. For each repetition of the sequence, the area of the repolarization transient is recorded. Phase cycling is used so that the transient area of the $-\pi/2$ sequence for each value of $\tau$ can be subtracted from the transient area of the $+\pi/2$ sequence so the echo signal decays towards a zero baseline. 

\begin{figure}[htbp]
\centering
\includegraphics[width=\columnwidth]{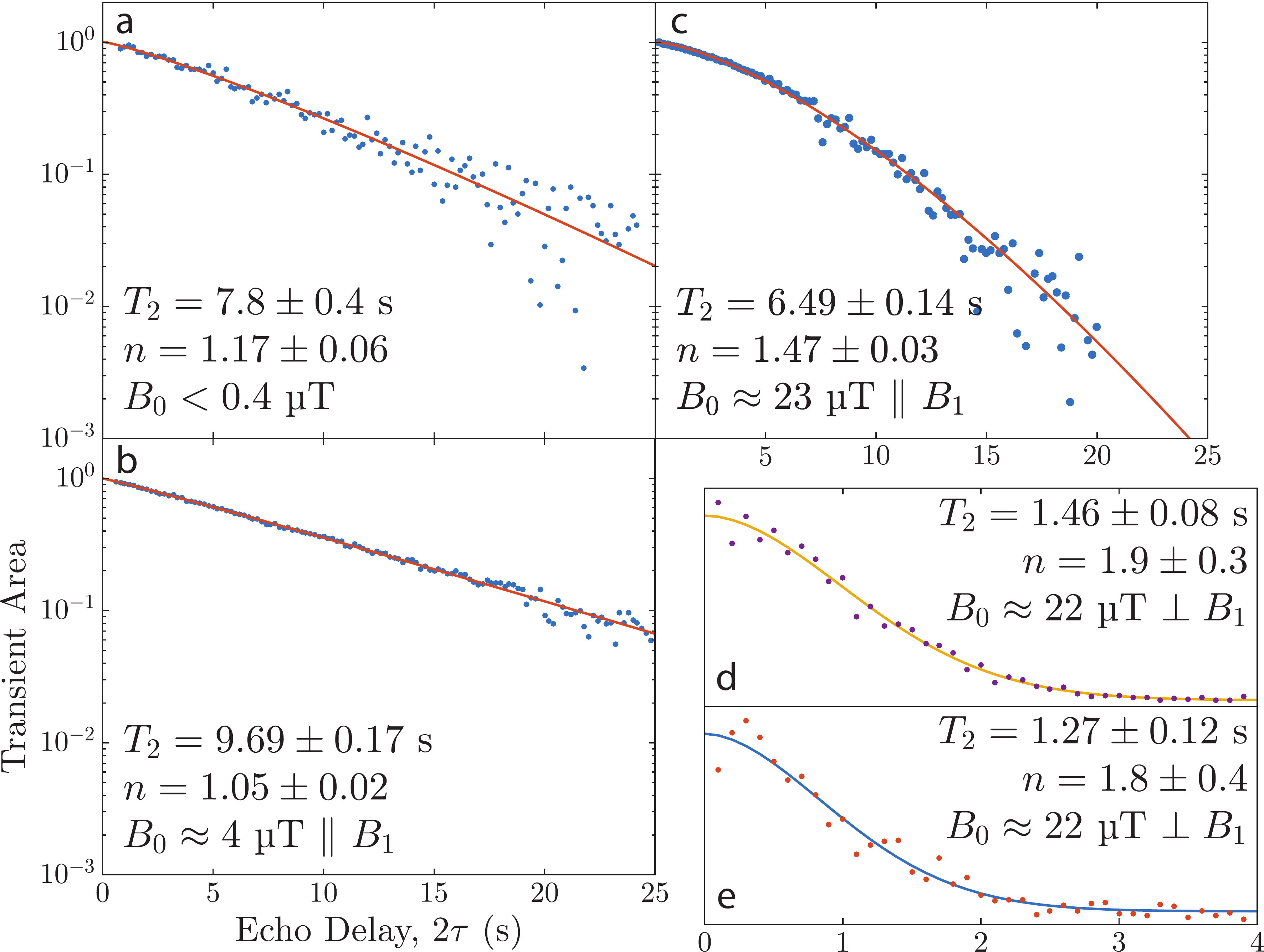}
\caption{\label{fig:all-hahn}Normalized Hahn echo data for the $S \to T_0$ (\textbf{a,b,c}) and $S \to T_\pm$ (\textbf{d,e}) transitions. All data was fit to a stretched exponential, $\exp\left(-(2\tau / T_2)^n\right)$, where $T_2$ is the decay constant and $n$ is the stretching parameter. 130\,mW of $9274.1888\,\text{cm}^{-1}$ excitation and 7\,\textmu{}W of $9550\,\text{cm}^{-1}$ (1047\,nm) above-gap excitation were used for initialization and readout. RF pulses were 0\,dBm for \textbf{a}, \textbf{b}, and \textbf{c} and 33\,dBm for \textbf{d} and \textbf{e}.
}
\end{figure}

Fig.~\ref{fig:all-hahn}a,b,c show Hahn echo decays for the $S \to T_0$ transitions at differing $B_0$. Although one would expect that the $T_2$ should be longest at the lowest field (Fig.~\ref{fig:all-hahn}a), since zero-field is the perfect clock transition for the $0\to0$ transition, we find that the $T_2$, with a small $B_0 \parallel B_1$ (Fig.~\ref{fig:all-hahn}b), is actually  longer. 

A possible explanation is that at lowest field, population in the $T_0$ state is mixing into the $T_{\pm}$ states since all three states are nearly degenerate. Once population has mixed into the $T_{\pm}$ states, it would dephase from the $T_0$ since the three \D{} states would have slightly different precession frequencies. This hypothesis is supported by the $n>1$ stretching parameter which suggests that spectral diffusion is limiting the $T_2$ decay.

A related explanation is that at lowest field there is an addressibility issue caused by the three $T$ states being nearly degenerate. In this case, the lower $T_2$ time would be due to our RF pulses partially driving the $S \to T_{\pm}$ transitions instead of selectively driving the $S \to T_0$ transition as intended.

Fig.~\ref{fig:all-hahn}c shows the Hahn echo decay of the $S \to T_0$ transition at a field of $23\,\text{\textmu{}T} \parallel B_1$. We see that $T_2$ at this larger field has decreased from the value observed for the smaller fields and the stretching parameter is the largest observed so far. At this field we are far enough from the clock transition that spectral diffusion related to magnetic field noise starts to affect $T_2$.

The final two plots, Fig.~\ref{fig:all-hahn}d and e, show the $S \to T_{\pm}$ transitions with a ``large'' $B_0$ field applied $\perp B_1$. For this configuration the Hahn echo data could only be collected by using maximum magnitude detection \cite{tyryshkin:2012}, since magnetic field noise causes the phase collected during the first half of the Hahn echo sequence to be imperfectly cancelled during the second half of the sequence.  This was to be expected, since, unlike the clock transition, these transitions have a linear frequency dependence on $B_0$.  Even though some coherence may remain at the end of the sequence, it has an arbitrary phase compared to the RF signal.  This is not a serious problem for inductively detected magnetic resonance, since both components of the remaining magnetization perpendicular to $B_1$ ($\hat{x}$ and $\hat{y}$) can be detected, enabling the coherence's magnitude to be extracted from a single ensemble measurement.  Since our readout is projective along the $\hat{z}$ axis, we collect many measurements at a given $\tau$ and use the largest absolute value \cite{Saeedi:2013}. 

\section{Conclusions}

Here we have demonstrated remarkably long coherence times for the hyperfine states of the phosphorus donor in silicon near zero magnetic field, and particularly for the $S \rightarrow T_0$ component, thanks to it being at a clock transition. This approach is easily extended to other shallow donors in silicon \cite{salvail:2015}, which will all have clock transitions at $B_0=0$.  The optical hyperpolarization and state readout at $\sim\!0$ magnetic field may offer major advantages for coupling donor spins to superconducting resonators \cite{Zollitsch:2015}.

\begin{acknowledgments}
This work was supported by the Natural Sciences and Engineering Research Council of Canada and by the Austrian Science Fund (FWF) through DK CoQuS. The \TwoEightSi{} samples used in this study were prepared from Avo28 crystal produced by the International Avogadro Coordination (IAC) Project (2004--2011) in cooperation among the BIPM, the INRIM (Italy), the IRMM (EU), the NMIA (Australia), the NMIJ (Japan), the NPL (UK), and the PTB (Germany).
\end{acknowledgments}


%

\end{document}